\documentstyle[12pt]{article}
\jot = 1.5ex\catcode`\@=11\renewcommand{\thefootnote}{\fnsymbol{footnote}}
\begin{document}\begin{titlepage}

\rightline{DFPD96/TH/29}

\rightline{\tt hep-th/9610026} 

\rightline{To be published in Phys. Rev. {\bf D}}

\vspace{1.4cm}

\centerline{\large{\bf SOLVING N=2 SYM BY REFLECTION SYMMETRY}} 

\vspace{0.4cm}

\centerline{\large{\bf OF QUANTUM VACUA\footnote[3]{Work supported by 
the European Commission TMR programme ERBFMRX--CT96--0045, to which 
M.M. and M.T. are associated.}}}

\vspace{1.3cm}

{\centerline{\sc Giulio
BONELLI, Marco MATONE, Mario TONIN}}

\vspace{0.8cm}

\centerline{\it Department of Physics ``G. Galilei'' - Istituto Nazionale di 
Fisica Nucleare}
\centerline{\it University of Padova}
\centerline{\it Via Marzolo, 8 - 35131 Padova, Italy}

\centerline{bonelli@padova.infn.it matone@padova.infn.it 
tonin@padova.infn.it}

\vspace{1.5cm}

\centerline{\bf ABSTRACT}

\vspace{0.6cm}

\noindent
The recently rigorously proved nonperturbative relation
$u=\pi i({\cal F}-a\partial_a{\cal F}/2)$, underlying $N=2$ SYM with 
gauge group $SU(2)$,
implies both the reflection symmetry $\overline{u(\tau)}=u(-\bar\tau)$
and $u(\tau+1)=-u(\tau)$ which hold exactly. The relation also implies 
that $\tau$ is the inverse of the uniformizing coordinate $u$ of the 
moduli space of quantum vacua ${\cal M}_{SU(2)}$, that is
$\tau:{\cal M}_{SU(2)}\to H$ where $H$ is the upper half plane. In this 
context, the 
above quantum symmetries are the key points to determine
${\cal M}_{SU(2)}$. It turns out that the 
functions $a(u)$ and $a_D(u)$, 
which we derive from first principles, actually coincide with the
solution proposed by Seiberg and Witten.
We also consider some relevant generalizations.

\end{titlepage}

\newpage

\setcounter{footnote}{0}

\renewcommand{\thefootnote}{\arabic{footnote}}

The exact results about $N=2$ SUSY Yang--Mills obtained by Seiberg
and Witten \cite{SW}
concern the low--energy Wilsonian effective action with at most two
 derivatives and four fermions. These terms are completely described by the
prepotential ${\cal F}$ whose instanton contributions
can be determined by recursion relations \cite{m}.
 In \cite{BM} it has been shown that the relation between 
${\cal F}$ and $u=\langle{\rm tr}\,\phi^2\rangle$ 
derived in \cite{m}, is connected to
the nonperturbative Renormalization Group Equation.
Related results concern the appearance of the WDVV equation
\cite{BM2,MMM} indicating that 
there are 
topological structures underlying $N=2$ SYM. In
\cite{BM2} it was argued 
that these aspects are connected
to associativity
which arises in considering divisors on moduli spaces
and therefore to quantum cohomology 
(see also \cite{Nekrasov}).

An interesting point is that 
Seiberg--Witten theory can be described 
in the framework of uniformization theory \cite{m,m2}
and the related Picard--Fuchs equations \cite{CerDaFeKLT}.
This aspect is also useful in considering the critical curve
 ${\cal C}=\{u|{\rm Im}\, a_D(u)/a(u)=0\}$ \cite{SW,m2,critical,BilalFerrari}.

In \cite{FinnellPouliot,FucitoTravaglini,DoreyKhozeMattis,HoweWest,Instanton}
nonperturbative investigations 
in the framework
of instanton theory and superconformal Ward identities
have been performed.
The aspects concerning integrability have been considered
in \cite{Integrable} whereas other related field
theoretical structures have been considered in \cite{QFT}
including some generalizations \cite{KLYTAF}.

All these results are a consequence of the Seiberg--Witten
derivation of the low--energy dynamics of $N=2$ SYM. 
One of the interesting aspects of the Seiberg--Witten results, 
both from a physical and mathematical
point of view, is that
their solution implies that all the instanton contributions
are determined by recursion relations \cite{m}. As a consequence,
the important problem of evaluating the relevant integrals
defining ${\cal F}_k$, has been done in a elegant way. We observe that
to understand the explicit structure of the integrals defining $k\ge 3$
is still an open problem which is of interest also for other QFT's.
These observations, while
indicating
 that the full consequences of the Seiberg--Witten results should
be further investigated, also make evident the necessity of proving 
what still remains at the conjectural level.
 Even if there is evidence supporting
 the Seiberg--Witten results, a clear proof is still lacking.

In this letter we will show that actually the basic structures
underlying 
$N=2$ SYM with gauge group $SU(2)$ are the asymptotic analysis 
which implies $u(\tau+1)=-u(\tau)$, a consequence
of the presence of the $\Theta$--angle, and 
the property of reflection symmetry $\overline u(\tau)=u(-\bar\tau)$
of the quantum vacua.
Of course, reflection symmetry is related to the CPT symmetry, 
which actually, together with the holomorphicity of the prepotential,
 turns out to be the crucial nonperturbative information.
 Therefore, what we will prove is that the
one--loop 
approximation and CPT arguments are sufficient to solve the theory
in the proper low--energy limit. Whereas the $T^2$ symmetry arises from 
the asymptotic analysis, the other generators of ${\Gamma(2)}$
turns out to be fixed by $T^2$ itself together with the reflection symmetry,
which can be seen as an alternative way to
define some subgroups of the discrete group $SL(2,{\bf Z})$.
Our method of characterizing some discrete groups from the symmetry 
properties of their fundamental domain is quite general and may
be also useful  
to shed new light not only in 
$N=2$ SYM but also in other quantum field theories.

An important point in our construction is that the  relation \cite{m}
\begin{equation}
u=\pi i \left[{\cal F}(a)-{a\over 2}
{\partial {\cal F}(a)\over \partial a}\right],
\label{a1}\end{equation}
has been proved
in the framework of multiinstanton calculations up to two instanton
contributions by Fucito and Travaglini 
 \cite{FucitoTravaglini} and at all orders
by Dorey, Khoze and Mattis 
\cite{DoreyKhozeMattis}. Furthermore, it has been derived in the framework
of superconformal Ward identities by Howe and West \cite{HoweWest}
proving also its generalization obtained in \cite{STYEY}.

In this letter, we will use the relation (\ref{a1})
in order to derive the uniformizing equation for the $u$--moduli 
space ${\cal M}_{SU(2)}$ of quantum vacua.
We will also show that two important consequences of the relation
are the reflection symmetry 
\begin{equation}
\overline{u(\tau)}=u(-\bar\tau),
\label{oij}\end{equation}
and
\begin{equation}
u(\tau-n)=(-1)^n u(\tau).
\label{1}\end{equation}
We stress the important point that as the relation
between $u$ and ${\cal F}$
has been derived both from multiinstanton calculations and
superconformal Ward Identities, it follows that we can exclude other 
unknown
nonperturbative effects besides the instanton contributions. 
As a consequence, both (\ref{oij}) and (\ref{1}) hold 
exactly. Eqs.(\ref{oij})(\ref{1}) 
turns out to be the key points to 
determine both ${\cal M}_{SU(2)}$ and its fundamental domain.
Indeed we shall prove from (\ref{oij}) and (\ref{1})
that ${\cal M}_{SU(2)}$ is the Riemann sphere with punctures
at $u=\infty$ and $u=\pm \Lambda^2$, the main conjecture in \cite{SW}.
In particular, it turns out that the 
functions $a(u)$ and $a_D(u)$, 
which we derive from first principles, actually coincide with those 
obtained by Seiberg and Witten.

Let us consider the chiral part of the low--energy effective
action for $N=2$ SYM with gauge group $SU(2)$. In $N=2$ superspace 
notation the part with at most two derivatives and
four fermions reads
\begin{equation}
{1\over 4\pi}{\rm Im}\,\int d^4x d^2\theta
d^2\tilde \theta {\cal F}(\Psi),
\label{oic}\end{equation}
where ${\cal F}$ is the prepotential and $\Psi$ the $N=2$ chiral 
superfield. The effective $\Theta$--angle and the gauge coupling constant
enter in the effective coupling constant $\tau=\partial_a^2{\cal F}(a)$ 
in the form
$$
\tau=\partial_a^2{\cal F}={\Theta\over 2\pi}+{4\pi i\over g^2}.
$$
The asymptotic expansion of ${\cal F}$ has the structure \cite{Seiberg}
\begin{equation}
{\cal F}=a^2\left[
{i\over \pi}\log {a\over \Lambda}+\sum_{k=0}^\infty {\cal F}_k
\left({a\over \Lambda}\right)^{-4k}\right],
\label{hfgty1}\end{equation}
where $\Lambda$ is the dynamically generated scale. Another important
result in \cite{Seiberg} is that actually at least
${\cal F}_1$ is non--vanishing.
For further purpose we write down the asymptotic expansion
of $\tau$
\begin{equation}
\tau={2i\over \pi}\log {a\over \Lambda}
+{3i\over\pi}+\sum_{k=0}^\infty {\cal F}_k
(1-4k)(2-4k)\left({a\over \Lambda}\right)^{-4k}.
\label{hfgty3}\end{equation}

By making some assumptions, 
Seiberg and Witten argued that the $u$--quantum moduli space is the 
thrice punctured Riemann sphere.
In particular, in \cite{SW} the exact 
form of the functions $a=a(u)$ and $a_D=a_D(u)=\partial_a{\cal F}$
 has been obtained.
This solution determines all the ${\cal F}_k$'s implicitly.

In \cite{m} it has been shown that the results in \cite{SW} imply
the relation (\ref{a1}).
This relation will be useful in determining both 
${\cal M}_{SU(2)}$
and the functions $a(u)$ and $a_D(u)$.
Eq.(\ref{a1}) has been checked 
in the framework of multiinstanton calculations up to two instanton
contributions in
 \cite{FucitoTravaglini} and at all orders
\cite{DoreyKhozeMattis}. Furthermore, it has ben derived in the framework
of superconformal Ward identities in \cite{HoweWest}.
The fact that (\ref{a1}) is rigorously proved is essential for our 
construction. 

By (\ref{hfgty1}) and (\ref{a1}) it follows that
the asymptotic expansion for $u={\cal G}(a)$ is 
\begin{equation}
{\cal G}(a)=a^2\sum_{k=0}^\infty {\cal G}_k \left({a\over\Lambda}\right)^{-4k},
\qquad {\cal G}_0={1\over 2},
\label{abas}\end{equation}
where ${\cal G}_k=2 \pi i k{\cal F}_k$. Furthermore, by instanton
theory \cite{FucitoTravaglini,DoreyKhozeMattis}
we have ${\rm Re}\,{\cal F}_k=0$.

Differentiating (\ref{a1})
with respect to $u$  we get 
$aa_D'-a_Da'={2i\over \pi}$.
This implies that
$a_D$ and $a$ are linearly independent 
solutions of a second--order linear differential 
equation with respect to $u$, that is
\begin{equation}
[\partial_u^2+V(u)]a=0=[\partial_u^2+V(u)]a_D,
\label{ll1}\end{equation}
for some unknown $V(u)=-a''/a=-a_D''/a_D$. 
Inverting (\ref{ll1}) 
we obtain a differential equation for $u$ with respect to $a$
\begin{equation}
\partial_a^2{\cal G}-a\left(\partial_a {\cal G}\right)^3
V\left({\cal G}\right)=0,
\label{odqj}\end{equation}
which implies recursion relations for 
${\cal G}_k=2 \pi i k{\cal F}_k$. The full 
Seiberg--Witten solution follows from (\ref{odqj}) once one proves that 
\begin{equation}
V(u)={1\over 4(u^2-\Lambda^4)},
\label{osiq}\end{equation}
so that Eq.(\ref{odqj}) becomes
\begin{equation}
\left(\Lambda^4-{\cal G}^2\right){\cal G}''+{1\over 4}a 
{{\cal G}'}^3=0,
\label{gdfet}\end{equation}
and by (\ref{abas}) \cite{m}
$$
{\cal G}_{n+1}={1\over 8{\cal G}_0^2(n+1)^2}\cdot
$$
\begin{equation}
\cdot\left\{
(2n-1)(4n-1){\cal G}_n
+2{\cal G}_0
\sum_{k=0}^{n-1}{\cal G}_{n-k}{\cal G}_{k+1}c(k,n)
-2\sum_{j=0}^{n-1}\sum_{k=0}^{j+1}{\cal G}_{n-j}
{\cal G}_{j+1-k}{\cal G}_{k}d(j,k,n)\right\},
\label{recursion2}\end{equation}
where $n\geq 0$, ${\cal G}_0=1/2$ and
$$
c(k,n)=2k(n-k-1)+n-1,
\qquad
d(j,k,n)=
[2(n-j)-1][2n-3j-1+2k(j-k+1)].
$$

Let us set 
\begin{equation}
T/2=V-V^{1/2}\partial_u^2\left(V^{-1/2}\right).
\label{tev}\end{equation}
{}From the identity
$$
V^{1/2}(u)\partial_u
\left[V^{-1}(u)\partial_u^2+1\right]=
\left[\partial_u^2 +{T(u)/2}\right]V^{-1/2}(u)\partial_u,
$$
and by (\ref{ll1}) we obtain 
\begin{equation}
\left[\partial_u^2+{T}(u)/2\right]V^{-1/2}(u) \partial_u a =0=
\left[\partial_u^2+{T}(u)/2\right]V^{-1/2}(u) \partial_u a_D,
\label{ll3}\end{equation}
where $T$ is the Schwarzian derivative (here $'\equiv \partial_u$)
$T(u)=\tau'''/\tau' -3(\tau''/\tau')^2/2$.

Since $\tau$ lives in the upper half plane $H$ (except that
at the possible singularities where ${\rm Im}\,\tau=0$), the 
polymorphic function
$\tau(u)=\partial_ua_D/\partial_ua$ 
may be seen as the inverse of the uniformizing coordinate
$u:H\to {\cal M}_{SU(2)}$. From the 
monodromy transformation properties of $\tau(u)$
we know that ${\cal M}_{SU(2)}\cong H/\Gamma$ where $\Gamma$ is the uniformizing
group to be determined.

We now observe that some information about the structure of 
${\cal M}_{SU(2)}$ already comes in considering the physical role of $u$.
For each
value of $u$, that is for each choice of representation of the vacuum
(which fixes the Hilbert space of states),
we should
determine the functions $a(u)$ and $a_D(u)$. Therefore, 
one has to consider the theory for each value 
of $u\in\widehat{\bf C}={\bf C}\cup\{\infty\}$.
In other words, defining ${\cal M}_{SU(2)}$ as the $u$--moduli
space means that $u$ is the uniformizing coordinate of
${\cal M}_{SU(2)}$  itself.
In this context, by  `singularities' we mean the values of $u$
which cause non trivial monodromies for $a(u)$ and $a_D(u)$. 
Geometrically this means that the $u$--space is the Riemann sphere 
$\widehat{\bf C}$ with 
$n$--punctures. 
Therefore, the unique non--trivial topological complications that we can 
expect are those induced from some particular values of $u$. In this 
context we observe that singularities imply symmetries. Actually, from the 
above discussion it follows that
$u(\gamma\cdot\tau)=u(\tau)$, $\gamma\in \Gamma$. That is, for any choice 
of $u$ there are infinitely many equivalent prepotentials \cite{m}
\begin{equation}
\gamma \cdot {\cal F}(a)=
{\cal F}(a)+{a_{11}a_{21}\over 2}a_D^2+
{a_{12}a_{22}\over 2}a^2+a_{12}a_{21}aa_D,
\qquad \left(\begin{array}{c}a_{11}\\a_{21}\end{array}
\begin{array}{cc}a_{12}\\a_{22}\end{array}\right)\in \Gamma.
\label{abbastanza}\end{equation}

In the following we will prove that the number of punctures is $3$ and will
determine $a(u)$ and $a_D(u)$ explicitly. As we will see we will do not need 
to make assumptions about the finiteness of the number of punctures.

Since the 
${\cal F}_k$'s are purely 
imaginary, it follows by (\ref{hfgty3}) that
$\overline{\tau(a)}=-\tau(\bar a)$, so that
$\overline{a(\tau)}=a(-\bar\tau)$ and by (\ref{abas})
we have the reflection symmetry (\ref{oij})
which is crucial for our construction.

Let us now consider the effect of the transformation $a\to e^{i\pi 
n/2}a$, $n\in{\bf Z}$, on ${\cal F}$, $\tau$ and $u$. By (\ref{hfgty1}) and 
(\ref{a1}) we have 
\begin{equation}
{\cal F}(a)\to e^{\pi i n}{\cal F}(a)-
e^{\pi i n}{n\over 2}a^2,\qquad \tau\to \tau-n,
\label{xwc}\end{equation}
and by (\ref{abas})
${\cal G}(e^{i\pi 
n/2}a)=(-1)^n {\cal G}(a)$ which is equivalent to
(\ref{1}).
We observe that
since from multiinstanton calculations both
${\cal F}_1$ and ${\cal F}_2$ are non--vanishing
\cite{Seiberg,FucitoTravaglini},
we can exclude 
that ${\cal G}(e^{i\pi 
n/m}a)\propto {\cal G}(a)$ for $m>2$.
We also stress that as 
a consequence of \cite{FucitoTravaglini,DoreyKhozeMattis,HoweWest}
 both (\ref{oij}) and (\ref{1}) hold exactly.

As Eq.(\ref{xwc}) shows, the  group elements acting on
$(a_D,a)$ have phases which do not appear
in the projective transformations of $\tau$.

By asymptotic analysis
we already know that there is a puncture at $u=\infty$.
We denote the other punctures and their image
in the closure of 
a given fundamental domain in $H$
by $u_k$ and $\tau_k=\tau(u_k)$, $k=0,\ldots,n-2$, respectively.
As well known from uniformization theory the $\tau_k$ correspond
to cusps on the real axis, the boundary of $H$. We fix the labelling of 
the punctures $u_0,\ldots,u_{n-2}$, in such a way that 
$\tau_{k+1}>\tau_k$, $k=0,\ldots,n-3$. Let us denote by $F$ the closure 
of the fundamental domain in $H$ which has non empty intersection
with the imaginary axis and by $\dot F$ its interior. 
By (\ref{1}) the width of $F$ is 2 whereas from the asymptotic behavior
$\tau\sim {i\over \pi} \log (2u/\Lambda^2)$, it follows that 
the $\tau$--image 
of the puncture at $u=\infty$ corresponds to the point at infinity.
This implies that the left and right parts of the boundary $\partial F$ 
of $F$ are 
two half--infinite vertical lines. We 
extend to $\forall k\in{\bf Z}$ the definition of the $\tau_k$'s by 
setting
\begin{equation}
\tau_{k+j(n-1)}=\tau_k+2j, \qquad j\in {\bf Z}.
\label{uhty}\end{equation}
We also set $\tau_0\le 0$ and $\tau_1>0$. 
Let $\tau^{(0)}$ be the image of the point $u=0$ in $F$ such that 
${\rm Re}\, \tau^{(0)}>0$. Observe that if 
$\tau^{(0)}\in \partial F$, then there is another point $\tau^{(-1)}$
in $\partial F$ such that $u(\tau^{(-1)})=0$. We require
${\rm Re}\, \tau^{(-1)}<0$, so that by construction
\begin{equation}
{\rm Re}\, \tau^{(0)}< {\rm Re}\, \tau^{(k)},\qquad \forall
\tau^{(k)}\in \{\tau|\tau\in G, u(\tau)=0, 
{\rm Re}\, \tau>0\},
\label{minore}\end{equation}
where $G=\cup_{k\in \bf Z}F^{(n)}$ and $F^{(n)}=\{\tau+2n|\tau\in F\}$.
One can check that by (\ref{oij}) and (\ref{1}) the above choices do not 
imply lost of generality.

We now start to determine both ${\cal M}_{SU(2)}$ and the fundamental 
domain. The starting point is to observe that
Eqs.(\ref{oij})(\ref{1}) in the $u=0$ case yield
\begin{equation}
0=u(\tau^{(0)})=u(\tau^{(0)}-1)=u(-\bar\tau^{(0)}).
\label{iuc}\end{equation}
We now show that by (\ref{iuc}) the point $u=0$ cannot be a puncture and 
that ${\rm Re}\, \tau^{(0)}=1/2$. Actually,
$\tau^{(0)}\notin \dot F$ otherwise either
$(\tau^{(0)}+1)\in \dot{F}$ 
or $(\tau^{(0)}-1)\in  \dot{F}$,
which is excluded by the one--to--one nature of the covering.
For the same reason $\tau^{(0)}$
is neither the image of a puncture nor belongs
to the half--infinite vertical lines in $\partial F$.
Hence, $\tau^{(0)}$ can be only on a half--circle
corresponding to the Poincar\'e geodesic connecting two cusps
$\tau_k$ and $\tau_{k+1}$ for some $k\in {\bf Z}$.
Furthermore, by (\ref{minore}) and (\ref{iuc})
we have
\begin{equation} 
{\rm Re}\tau^{(0)}={1\over 2}.
\label{dihq}\end{equation}

Let us denote by ${\cal R}$ and ${\cal I}$ the loci
${\rm Im}\, u=0$ and ${\rm Re}\, u=0$ respectively.
We also set $\tau_y={\rm Im}\, \tau$. In order to 
determine the image of ${\cal R}$ in $G$ we observe that reflection 
symmetry implies ${\rm Im}\, u(\tau=i\tau_y)=0$ that by (\ref{1}) 
extends to ${\rm Im}\, u(\tau=k+i\tau_y)=0$, $k\in {\bf Z}$.
A similar reasoning yields ${\rm Re}\, 
u(\tau=k+1/2+i\tau_y)=0$, $k\in {\bf Z}$. Summarizing, we have
\begin{equation}
{\rm Im}\,u(\tau=k+i\tau_y)=0, \qquad
{\rm Re}\, u(\tau=k+1/2+i\tau_y)=0,\qquad k\in {\bf Z}.
\label{dojqds}\end{equation} 

{}From the above investigation it follows that the points $\tau(u=0)$ in 
$G$ correspond to the end--points of the vertical lines belonging
to the $\tau$--image of the ${\cal I}$ locus. Let us denote
by $V_{k+1/2}$, $k\in{\bf Z}$ these vertical lines.
Since the holomorphicity of ${\cal F}(a)$ \cite{Gates,Seiberg}
implies the holomorphicity
of $\tau(u)$, it follows that 
the angles on the $u$--space are preserved
on the fundamental domains except that at the possible punctures.
As $u=0$ is not a puncture,
for each $k\in {\bf Z}$, the line $V_{k+1/2}$ is perpendicular 
to the curve 
image of ${\cal R}$ in $G$ at the point
$\tau^{(0)}+k$.
We note that the locus $V\in\dot{F}$ corresponding to points 
in the $\tau$--image of ${\cal R}$ is the intersection of the 
imaginary axis with $\dot F$, that is 
$$
V=\dot F\cap \{\tau| {\rm Re}\, \tau=0\}.
$$
 It follows that the only 
possibility for the $\tau$--image of ${\cal R}$ to be a line in $F$ which is 
continuous, except that at the cusps, is that the full boundary
$\partial F$ itself be in the $\tau$--image of ${\cal R}$.
Therefore, we have
\begin{equation}
\tau:{\cal R} \to \partial F\cup V.
\label{realinf}\end{equation}
Furthermore, since $\tau_k\in \partial F$, it follows that the punctures 
are real
\begin{equation}
{\rm Im}\,u_k=0.
\label{iupo}\end{equation}
By reflection symmetry and (\ref{iupo}) we have 
$u_k=u(\tau_k)=u(-\tau_k)$, and by (\ref{1})
$\tau_k\in {\bf Z}$. 
Since $\tau_0\le 0$, $\tau_1>0$ and $\tau_{k+1}>\tau_k$, we have
\begin{equation}
\tau_k=k, \qquad k\in {\bf Z}.
\label{risultato}\end{equation}
It follows that ${\cal M}_{SU(2)}$ is the Riemann sphere with
punctures
at $u_0=u(\tau=0)$, $u_1=u(\tau=-1)=u(\tau=1)=-u_0$ and $\infty$.
As well known this surface is uniformized by $\Gamma(2)$. In order to 
find the value of $u_0$ and the explicit form of $a(u)$ and $a_D(u)$,
we follow \cite{m} by first
considering the explicit expression of the projective connection
$T(u)=\{\tau, u\}$. For a $n$--punctured 
Riemann sphere with a puncture at infinity we have 
(see for example \cite{Nehari})
\begin{equation}
T(u)=
\sum_{i=0}^{n-2}\left[{1\over 2(u-u_i)^2}+
{c_i\over u-u_i}\right].
\label{7}\end{equation}
where
the $c_i$'s, called accessory parameters, satisfy the constraints
\begin{equation}
\sum_{i=0}^{n-2}c_i=0,\qquad \sum_{i=0}^{n-2}c_iu_i=1-{n\over 2}.
\label{8b}\end{equation}
In our case $n=3$, so that $c_1=-c_0=1/4u_0$, and Eq.(\ref{ll3}) becomes
\begin{equation}
\left[\partial_u^2 +{3u_0^2+u^2\over 4(u_0^2-u^2)^2}\right]\psi=0.
\label{11}\end{equation}
Note that by (\ref{tev}) we have $V(u)=1/4(u^2-u_0^2)$. To find 
$u_0$, $a(u)$ and $a_D(u)$ we first note that (\ref{11}) is solved
by the Legendre functions $P_{-1/2}$ and $Q_{-1/2}$. This fact and the 
asymptotic analysis imply \cite{m}
\begin{equation}
a_D(u,\Lambda)
={\sqrt 2\over \pi}\int_{\Lambda^2}^u {dx \sqrt{x-u}\over 
\sqrt{x^2-\Lambda^4}},\qquad
a(u,\Lambda)={\sqrt 2\over \pi}\int_{-\Lambda^2}^{\Lambda^2}
 {dx \sqrt{x-u}\over 
\sqrt{x^2-{\Lambda^4}}},
\label{14}\end{equation}
which actually coincides with the solution proposed by Seiberg and Witten 
\cite{SW}. Observe that $u_0=u(\tau=0)=\Lambda^2$.
We also note that as a consequence of the nonperturbative 
quantum symmetries (\ref{oij}) and (\ref{1}) of $N=2$ SYM, although the 
width of $F$ is 2, the tessellation of $H$ by $\Gamma(2)$ has an 
automorphism under the shift $\tau\to \tau+1$. In this context we observe 
that the asymptotic behavior
fixes the sign ambiguity.
In particular,
by (\ref{14}), whose normalization is fixed by (\ref{hfgty3}) 
and (\ref{abas}),
it follows that the positive imaginary axis of $H$ is in the 
$\tau$--image of the 
real points $u>\Lambda^2$. 

We observe that in our construction we did not make any assumptions about 
finiteness of the number of punctures.

Let us now shortly consider how the approach works in other cases.
As the relation between $u$ and the prepotential
is the crucial point, we have to see if there is an extension of it. 
In the $SU(2)$ case with $N_f\ne 0$ the relation has a very simple 
generalization.
In the case of rank group $r$ the relation has the form \cite{STYEY}
\begin{equation}
u={1\over 4\pi b_1}\left({\cal F} -
{1\over 2}\sum_{k=1}^ra_k{\partial {\cal F}\over \partial a_k}\right),
\label{vari}\end{equation}
where $b_1$ is the one--loop contribution to the beta--function.
Eq.(\ref{vari}) has been proved in \cite{HoweWest}.
Of course, besides $u$ there are other $(r-1)$--moduli coordinates.
In the $SU(3)$ case, the relation between 
$v=\langle {\rm tr}\,\phi^3\rangle$ and ${\cal F}$ has been derived
in \cite{BM2}.
Other informations concerning the
structure of the theory may be obtained from the asymptotic analysis.
As in the $SU(2)$ case, this asymptotic fixes the monodromy at infinity.
Then again, using asymptotic analysis and
CPT arguments (reflection symmetry), will uniquely fix the structure of the
fundamental domain for the monodromy group. The fact that the situation still 
involves uniformization, follows by the geometrical analysis 
considered in \cite{BM2}
where it was proved that
$$
{\cal M}_{SU(3)}\cong{\cal S}/\Gamma,
$$
with ${\cal S}$ the genus two $\tau_{ij}$--space, a subvariety
of the genus 2 Siegel upper--half space of complex codimension 1 
covering ${\cal M}_{SU(3)}$.
Therefore, the uniformization arguments still work in the higher rank 
case.
Actually, this was also of mathematical interest as considers the problem
of formulating the uniformization theorem in the case of moduli space of 
Riemann surfaces rather than the Riemann surfaces themselves. 
However the uniformization
 in the case of the quantum moduli space seems to be easier than the 
general case as the kind
of Riemann surfaces described by points in ${\cal M}_{SU(n)}$ actually
have a higher symmetry coming just from the asymptotic analysis
and from CPT arguments. This reflects in the fact that the position
 of the branching points in the Riemann sphere is highly symmetric. 
This is equivalent to peculiar kinds of Riemann period 
matrices
which should reflect in some interesting number theoretical properties.

It is a pleasure to thank 
Frank Ferrari, Francesco Fucito, Pier Alberto Marchetti, 
Paolo Pasti and Gabriele Travaglini for discussions.

\end{document}